\documentstyle[twoside,fleqn,espcrc2,epsfig]{article}
\def\be{\begin{equation}}
\def\ee{\end{equation}}
\def\beq{\begin{eqnarray}}
\def\eeq{\end{eqnarray}}

\begin{document}

\title{SUSY CONTRIBUTIONS TO $R_b$ AND TOP DECAY}

\author{D.P. Roy \address{Tata Institute of Fundamental
Research, Bombay 400 005, India}}
       
\begin{abstract}
I report on a systematic analysis of the MSSM parameter space to
obtain the best SUSY solution to the $R_b$ anomaly within the
constraint of top quark decay. Phenomenological implications for
top decay and direct stop production at the Tevatron collider
are discussed.
\end{abstract}

\maketitle

The LEP value of $R_b = .2202 \pm .0016$, obtained by assuming
the SM value for $R_c$; shows a 2.8$\sigma$ excess over the SM
prediction of $R_b = .2157$ \cite{one}. While it is not possible
to explain such a large excess via SUSY, we shall consider a
SUSY contribution of half this magnitude, i.e.
\be
\delta R_b = .0022 \pm .0004 ,
\ee
as a viable solution to the $R_b$ anomaly \cite{two}. This will
bring the theoretical value of $R_b$ within 1.6$\sigma$ (90\%
CL) of the LEP data.  Moreover the resulting drop in $\alpha_s
(M_Z)$,
\be
\delta \alpha_s (M_Z) \simeq -4 \delta R_b \simeq - .01 ,
\ee
will bring its estimate from $\Gamma_Z^{\rm had}$ in agreement
with the DIS value \cite{three}. We shall assume an optimistic
upper limit for the SUSY BR of top decay \cite{two},
\be
B_s < 0.4 ,
\ee
whose validity will be discussed below.

In the low $\tan \beta~(\simeq 1)$ region of our interest, the
SUSY contribution to $R_b$ comes from the lighter stop
\be
\tilde t_1 = \cos \theta_{\tilde t} \tilde t_R - \sin \theta_{\tilde
t} \tilde t_L ,
\ee
and the charginos
\be
\tilde W_{iL} = V_{i1} \tilde W^\pm_L + V_{i2} \tilde H^\pm_L \ \:, \
\: \tilde W_{iR} = U_{i1} \tilde W^\pm_R + U_{i2} \tilde H^\pm_R . 
\ee
The dominant SUSY contributions to $Z \rightarrow \bar b b$ come
from the triangle diagrams involving lighter stop and chargino
exchanges, $\tilde W_1 \tilde W_1 \tilde t_1$ and $\tilde t_1
\tilde t_1 \tilde W_1$ \cite{four,five}. The $b$ vertices,
common to both the diagrams, are dominated by the $b_L \tilde
W_{1L} \tilde t_{1R}$ Yukawa coupling
\be
\Lambda^L_{11} \sim - m_t \cos \theta_{\tilde t} V_{12}/\sqrt{2} M_W
\sin \beta .
\ee
The $Z$ vertex for the former process is determined by the
couplings
\be
O^L_{11} \sim \cos 2 \theta_W + U^2_{11} \ , \ O^R_{11} \sim
\cos 2
\theta_W + V^2_{11} , 
\ee
while it is suppressed for the latter by the U(1) coupling
factor $\sim \sin^2 \theta_W$.  Thus the large $b$ vertex (6)
favours large higgsino component of $\tilde W_1$ corresponding
to the higgsino dominated region ($| \mu | \ll M_2$); but the
large $Z$ vertex (7) favours large gaugino components of $\tilde
W_1$ corresponding to the gaugino dominated region ($| \mu | \gg
M_2$) \cite{five}. As we shall see below, the combined
requirements of large $b$ and $Z$ vertices give the best value
of $\delta R_b$ for the mixed region ($| \mu | \simeq M_2$)
corresponding to a $\tilde \gamma$ dominated LSP, rather than
the higgsino dominated region favoured by some earlier works
\cite{two}.

The SUSY BR for top decay ($t \rightarrow \tilde t_1 \tilde
Z_{1,2}$) is governed by the Yukawa couplings of the lighter
neutralinos
\be
\tilde Z_i = N_{i1} \tilde B + N_{i2} \tilde Z + N_{i3} \tilde H_1 +
N_{i4} \tilde H_2 \ ,
\ee
{\it i.e.}
\be
C_i^{L(R)} \sim m_t N_{i4} \cos \theta_{\tilde t} (\sin
\theta_{\tilde t})/M_W \sin \beta \ .
\ee
Thus the higgsino dominated region, corresponding to $\tilde
H_2$ dominated $\tilde Z_1$ and $\tilde Z_2$, leads to large
values of $B_s$. One gets relatively small $B_s$ in the mixed
region, where only $\tilde Z_2$ is higgsino dominated. Thus the
top decay constraint (3) favours the mixed region over the
higgsino dominated one as well.

Figure 1 shows the SUSY contributions to $R_b (\delta R_b)$ and
top BR ($B_s$) at $\tan \beta = 1.1$ for 3 representative points
in the $M_2,
\mu$ plane \cite{six}, {\it i.e.} 
\be
M_2, \mu = (a) 150, -40 \ \ (b) 60, -60 \ \ (c) 40, -70 {\rm
GeV}.
\ee
The higgsino dominated region ($a$) is seen to give too small a
value of $\delta R_b~(\leq .0014)$ for $B_s < 0.4$. On the other
hand one gets acceptable solutions to $\delta R_b$ (1) for $B_s
< 0.4$ in the mixed region, represented by the points ($b$) and
($c$). The best solutions to $\delta R_b$ and $B_s$ are obtained
for
\be
m_{\tilde t_1} \simeq 60~{\rm GeV} \ \ {\rm and} \ \
\theta_{\tilde t}
\simeq -15^\circ , 
\ee
where the stop mass is below the D$0\!\!\!\!/$ excluded region
$m_{\tilde t_1} \neq 65 - 88$ GeV \cite{seven}. However, the
point (c) also gives acceptable values of $\delta R_b$ for a
relatively large stop mass of 90-100 GeV.

Figure 2 shows the contour plots of $\delta R_b$ and $B_s$ in
the $M_2, \mu$ plane for $\tan \beta = 1.1$ and 1.4, with the
optimal choices of stop mass and mixing (11). The points
($a,b,c$), shown as bullets, are chosen close to the LEP
boundary so as to give the best values of $\delta R_b$ in their
respective regions. The higgsino dominated region, represented
by the point $a$, clearly corresponds to a low $\delta R_b$
along with an excessively large $B_s$. One sees a 30-40 \%
increase in $\delta R_b$ along with a similar fall in $B_s$ as
one goes down to the mixed region, represented by the points $b$
and $c$. Consequently one gets acceptable values of $\delta R_b$
(1) for $B_s$ = 0.3 -- 0.4 in this region. By far the best
solution to $\delta R_b$ and $B_s$ is offered by the point $c$.
But it corresponds to $M_{\tilde g} \simeq$ 160 GeV, ($M_{\tilde
W_1} \simeq$ 80 GeV), which is just above the Tevatron gluino
mass limit of $M_{\tilde g} \geq$ 150 GeV \cite{eight},
represented by the x-axis. On the other hand the point $b$
corresponds to $M_{\tilde g} \simeq$ 240 GeV and $M_{\tilde W_1}
\simeq$ 95 GeV, which are safely above the reaches of Tevatron
and LEP-2.

The SUSY BR of
\be
B_s = 0.3 - 0.4
\ee
has phenomenological implications for the $\bar t t$ events at
Tevatron \cite{nine,ten}. The isolated lepton plus $n$-jet
events with $b$-tag \cite{nine} come from the SM decay of one
top ($\bar t \rightarrow \bar b \ell\nu$), while the other
undergoes SM or SUSY decay
\be
t \rightarrow bW \rightarrow b q \bar q^\prime ,
\ee
\be
t \rightarrow \tilde t_1 \tilde Z_1 \rightarrow \tilde Z_1 c
\tilde Z_1 q \bar q.
\ee
The SUSY decay is characterised by a lower detection efficiency
(due to the absence of lepton and $b$ and fewer visible jets,
but a larger missing-E$_T$ ($E_T\!\!\!\!\!\!/$).  They lead to
the following differences with respect to the SM prediction
\cite{six}.
\begin{enumerate}
\item[{(i)}]
There is a shift of $\sim$ 10\% (i.e.3--4) of the above $\bar t t$
events from $\geq$ 4 jets to the 2 jets channel. Such a shift
seems to be favoured by the preliminary CDF data \cite{nine},
but with large uncertainty.
\item[{(ii)}]
The detection efficiency and hence the experimental
cross-section in the $\geq$ 3 jets channel are reduced by a
factor
\be
(1-B_s) (1-B_s/3) = 2/3 - 1/2 .
\ee
This is disfavoured by the CDF cross-section.
\be
\sigma_t = 7.5 \pm 1.8~{\rm pb}, m_t = 175.6 \pm 9~{\rm GeV},
\ee
which is already larger than $\sigma_t^{\rm QCD} (175) \simeq
5.5$ pb
\cite{eleven}. But the 1.6$\sigma$ (90\% CL) lower limits of
$\sigma_t$ and $m_t$ would correspond to an experimental
$\sigma_t (= 4.5$ pb) of $\sim$ 1/2 the size of the
corresponding $\sigma^{\rm QCD}_t (160) \simeq 9$ pb, as
required by (14). This is the basis for the $B_s$ limit (3). It
will be easier to satisfy with the D$0\!\!\!\!/$ cross-section,
$\sigma_t = 5.3 \pm 1.6$ pb \cite{ten}.
\item[{(iii)}]
There is an 50\% enhancement of the large transverse mass $(M_T
(\ell E_T\!\!\!\!\!\!/) > $ 120 GeV) tail of the above $\bar t
t$ events.  Similarly one expects a $\sim$ 25\% deficit in the
$\bar t t$ events in the dilepton and double $b$-tagged
channels. Although these are $\leq 1\sigma$ effects for the
current Tevatron luminosity of $\sim$ 100 pb$^{-1}$, they will
be 2-3$\sigma$ effects for the $\sim -1~{\rm fb}^{-1}$
luminosity expected with the main injector run.
\end{enumerate}

Finally, the large stop mass (90--100 GeV) solution for the
point $c$ would correspond to the charged current decay $\tilde
t_1 \rightarrow b \tilde W_1$. This would lead to a detectable
dilepton signal from stop pair production at Tevatron.

The work reported here was done in collaboration with M. Drees,
R.M.  Godbole, M. Guchait and S. Raychaudhuri \cite{six}.

\newpage
\begin{enumerate}
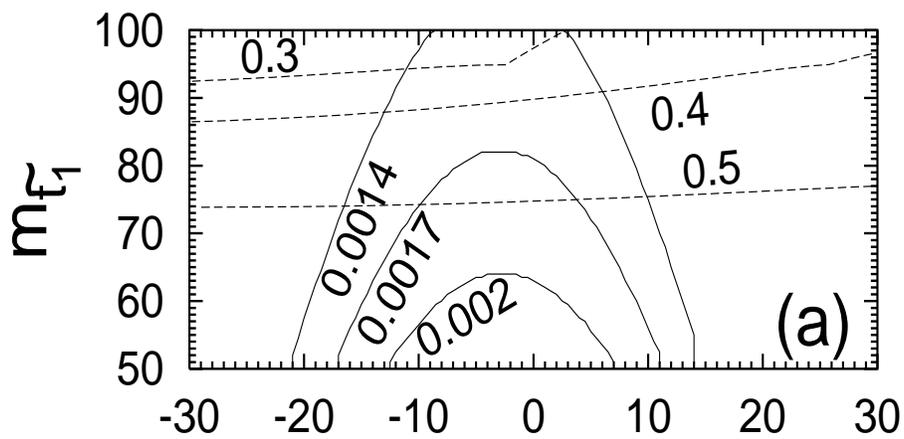
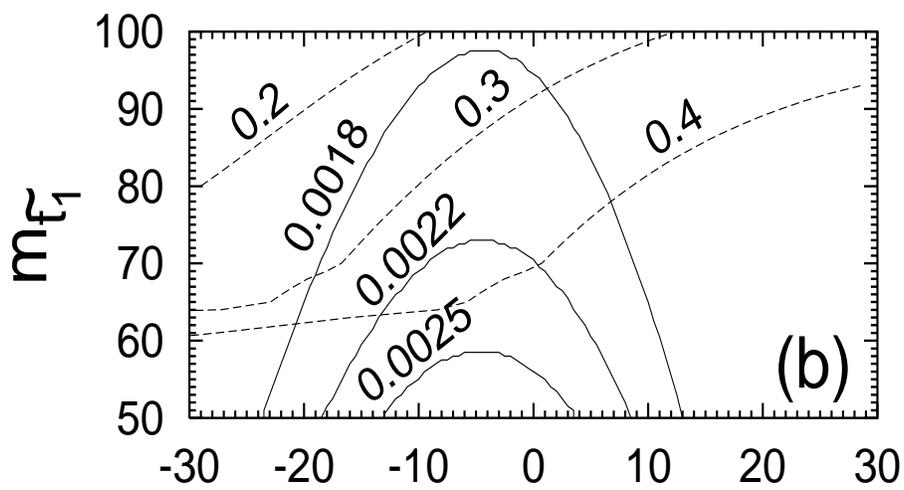
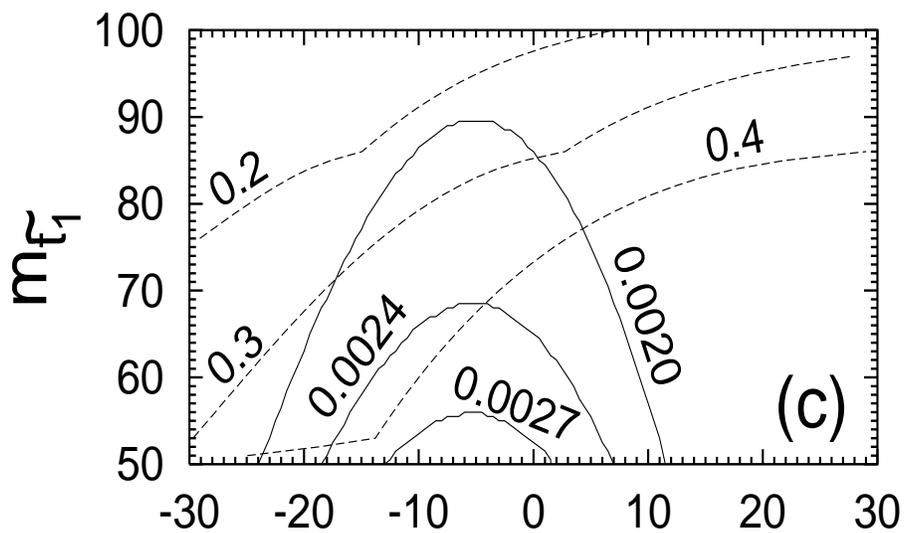
\item[{\rm Fig. 1.}] SUSY contributions to $R_b$ (solid) and the
top BR (dashed) are shown as contour plots in stop mass and
mixing angle for $M_2,\mu = (a) 150, -40 (b) 60, -60, (c)
40,-70$ GeV with $\tan(\beta) = 1.1$.
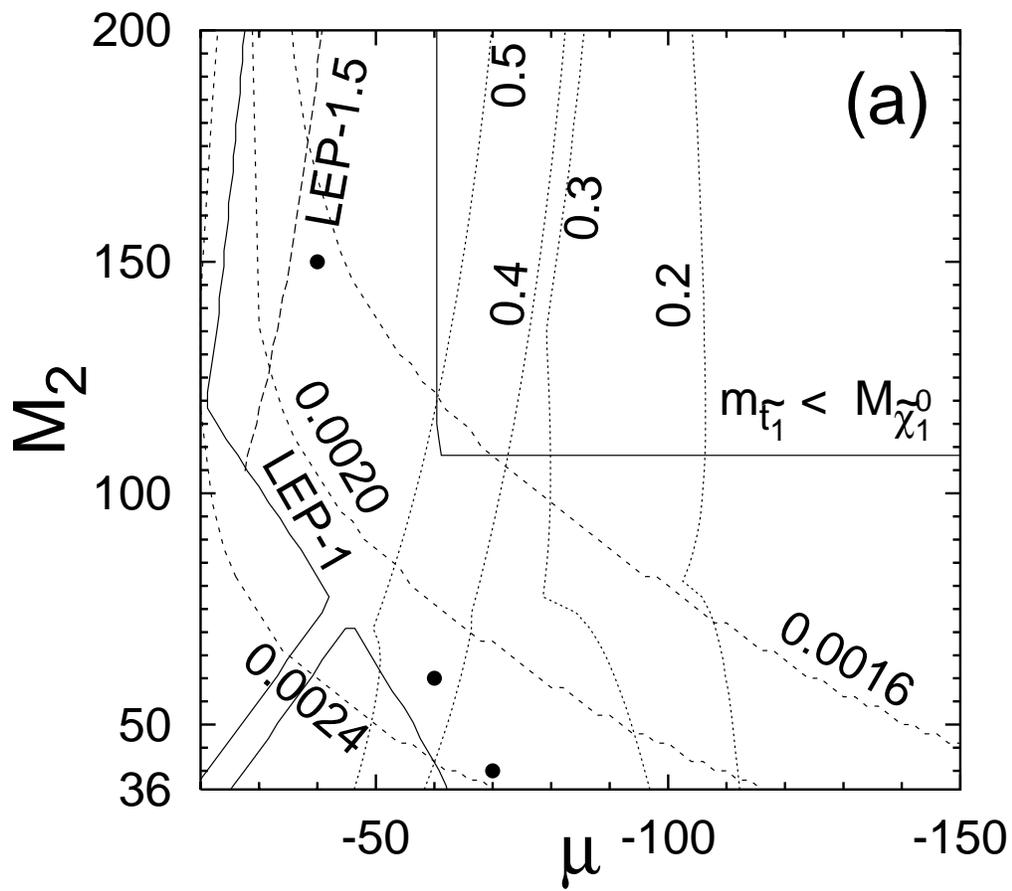
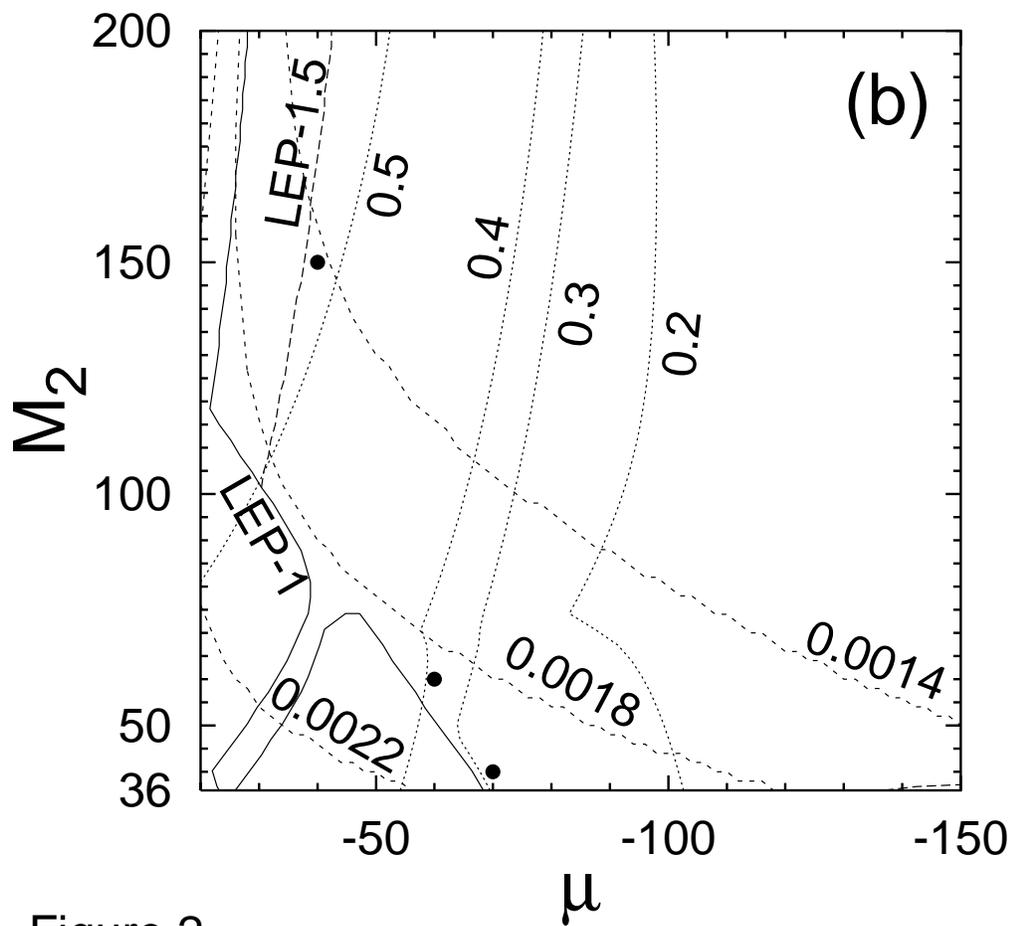
\item[{\rm Fig. 2.}] SUSY contributions to $R_b$ (dashed) and
top BR (dotted) shown as contour plots in the $M_2, \mu$ plane
for stop mass (mixing angle) of 60 GeV (-15$^\circ$), with $\tan
\beta$ = ($a$) 1.1, ($b$) 1.4. The boundary of the region
$m_{\tilde t_1} < M_{\tilde Z_1}$ is not shown in ($b$).
\end{enumerate}

\end{document}